\documentclass[twocolumn,superscriptaddress]{revtex4-1}
\usepackage{amsmath, bm, physics,amssymb}
\usepackage{color}
\usepackage{graphicx}
\usepackage{siunitx}
\usepackage{txfonts}

\usepackage[whole]{bxcjkjatype} 
\usepackage[colorlinks=true,urlcolor=blue,citecolor=blue,linkcolor=blue,breaklinks=true]{hyperref}
\input glyphtounicode
\newcommand{\smrm}[1]{_{\mathrm{#1}}}
\newcommand{\uprm}[1]{^{\mathrm{#1}}}

\begin{document}
\title{Emergent induction in magnetic Weyl semimetals}

\author{Takahiro Anan}
\affiliation{Department of Applied Physics, The University of Tokyo, Hongo, Tokyo,
  113-8656, Japan}
\affiliation{Department of Physics, Kyoto University, Kyoto,
  606-8502, Japan}  
\author{Takahiro Morimoto}
\affiliation{Department of Physics, Kyoto University, Kyoto,
  606-8502, Japan}
\date{\today}

\begin{abstract}
We theoretically study emergent electromagnetic responses in Weyl semimetals.
Focusing on magnetic Weyl semimetals, we develop a general theory of emergent induction driven by magnetic dynamics. 
We show that magnetoelectric (ME) responses in Weyl semimetals give rise to emergent induction mediated by magnetization dynamics. 
Using effective two-band models for magnetic Weyl semimetals, we derive a formula for the ME response that includes both intraband and interband contributions. 
The resulting formula shows that the intraband contribution is proportional to the relaxation time $\tau$, whereas the interband contribution is associated with the separation of the Weyl nodes. 
Applying the general formula to a model of polar Weyl ferromagnets, we demonstrate that the dynamics of the toroidal moment is closely related to the emergent inductive response in polar Weyl ferromagnets, as recently discovered by Suzuki et al. [Y. Suzuki et al. arXiv:2607.12322]. 
The chemical-potential dependence of the inductance indicates that the emergent electromagnetic response is enhanced in the energy range of the Weyl dispersion, reflecting the topological nature of Weyl semimetals.
\end{abstract}

\maketitle

\section{Introduction}
Weyl semimetals are topological phases of matter where the low-energy excitations are described by Weyl fermions with gapless linear dispersion~\cite{Murakami07,Armitage,Xu2015}.
The topological nature of the Weyl fermion gives rise to a variety of anomalous transport phenomena, including Hall responses, and is accompanied by topological surface states known as Fermi arcs~\cite{Wan11,Yang2011,Zyuzin2012,Burkov2014}. 
In magnetic Weyl semimetals, the exchange coupling between itinerant electrons and localized magnetic moments further enriches this physics since the positions and structures of Weyl points become sensitive to the magnetic configuration~\cite{Armitage,bernevig2022progress,nagaosa2020transport,ozawa2026magnetic,ArakiRev2020}.
This phenomenon is intuitively understood by assuming a Weyl fermion with full spin-momentum locking, described by a Hamiltonian of a Weyl fermion with $H=v\bm{k} \cdot \bm{\sigma}$ with spin Pauli matrices $\bm{\sigma}$. Once we introduce coupling to the magnetization $\bm{m}$, the Hamiltonian reads 
\begin{align}
  \mathcal{H}=v\bm{k} \cdot \bm{\sigma}+\bm{m}(t)\cdot \bm{\sigma}. 
  \label{eq:Weyl}
\end{align}
The time-dependent magnetization $\bm{m}(t)$ effectively behaves as an emergent vector potential $\bm{a}_{\mathrm{em}}(t)=\frac{\hbar}{ev}\bm{m}(t)$ which is associated with the shift of the Weyl point in momentum space.
Consequently, time-dependent spin dynamics can shift the Weyl points in momentum space and generate emergent electric fields $\bm{e}_{\mathrm{em}}\equiv -\partial_t \bm{a}_{\mathrm{em}}=-\frac{\hbar}{ev}\partial_t \bm{m}(t)$~\cite{Araki2018,Semenov2023,Harada2023}.
While a general Weyl fermion contains a mixture of spin and orbital degrees of freedom~\cite{Ozawa2024}, this mechanism provides a natural route to strong magnetoelectric responses in magnetic Weyl systems.

A related magnetism-mediated dynamical transport phenomenon is emergent inductance, in which the dynamics of magnetic degrees of freedom mediate an electrical response.
When an electric current or electric field drives spin dynamics, the delayed response of the localized moments feeds back into charge transport and produces a complex impedance~\cite{Nagaosa2019,Kurebayashi2021,Ieda2021,Yamane2022,Araki2023,Oh2024,Anan2025,Yamada2026,Araki2026,Wang2026,Yokouchi2020,Kitaori2021,Kitaori2023,Kitaori2024,Matsushima2024,Zhang2025,Yamada2026}.
Current-induced magnetization responses associated with spin--orbit torques have already been observed in Weyl-semimetal systems~\cite{Li2018,Tang2021,Bainsla2024}, and the corresponding ME responses have been studied theoretically in magnetic Weyl semimetals~\cite{Kurebayashi2019,Kurebayashi2021b,Meguro2025}.
Moreover, a recent experiment on a polar Weyl ferromagnet PrAlGe has revealed that the magentization dynamics driven by the applied electric current gives rise to an emergent Hall inductance through the dynamics of the toroidal moment ($T=P\times M$ with polarization $P$ and magnetization $M$), which is named ``emergent toroidal induction''~\cite{Suzuki26}.

In this work, we develop a theory of emergent inductance in Weyl semimetals based on these ME responses.
We derive the magnetoelectric (ME) response in magnetic Weyl semimetals by including both Fermi-surface and interband contributions, and the resulting inductance mediated by localized-spin dynamics. 
We show that the ME response consists of two distinct contributions: a dissipative Fermi-surface contribution proportional to the relaxation time $\tau$ and to the Fermi-surface area, and an interband contribution associated with the Weyl-node separation, or equivalently with the length of the Fermi arc. 
These Weyl-specific ME effects enhance the emergent inductance.

\section{Magnetoelectric effect of Weyl fermions}
In this section, we study magnetoelectric (ME) effect of Weyl fermions.
In the ME effect, the applied electric field induces spin magnetization or the spin dynamics induces the electric current. 
Previous theories have established a dissipative intraband response~\cite{Kurebayashi2019,Kurebayashi2021b} and a nondissipative interband response in a numerical calculation~\cite{Kurebayashi2019,Meguro2025}.
Here, we derive analytical expressions for both the intraband and interband contributions using two-band effective models for Weyl fermions, which are then used to formulate the emergent inductance.

\subsection{Intraband contribution to the ME effect}
To study the intraband contribution to the ME effect, we reproduce the dissipative response discussed in Ref.~\cite{Kurebayashi2021b} using Boltzmann transport theory for a simple two-band model of a Weyl fermion.
The effective Hamiltonian near an untilted Weyl point under a perturbative external magnetic field due to the Hund coupling with localized spins is given by, 
\begin{align}
  \mathcal{H}=\bm{k}^\mathrm{T}K\bm{\tau}+\bm{S}^\mathrm{T}(t)M\bm{\tau}, \label{eq:WeylGeneral}
\end{align}
where $\bm{S}(t)$ represents time-dependent fluctuations of classical localized spins. 
The Pauli matrices $\tau$ act on the two dimensional Hilbert space consisting of a mixture of spin and orbitals generally. 
Accordingly, $K$ is a three-by-three matrix that determines the momentum-space structure and anisotropic Fermi velocities of the Weyl point, and $M$ is a three-by-three matrix that determines the spin component of the two dimensional Hilbert space for the Weyl fermion and describes the strength of the Hund coupling between itinerant electrons and localized spins.

To derive the intraband contribution to the ME effect, we consider the Boltzmann equation,
\begin{align}
  -\frac{f-f_0}{\tau}=\frac{df}{dt}=\frac{df}{d \xi_{\bm{k},u}}\left(\frac{\partial \xi_{\bm{k},u}}{\partial \bm{k}}\cdot \dot{\bm{k}}+\frac{\partial \xi_{\bm{k},u}}{\partial t}\right), \label{eq:BoltzmannEq}
\end{align}
for the distribution function $f$, where $\xi_{\bm{k},u}\equiv \epsilon_{\bm{k},u}-\mu$.
Here, $\mu$ is the chemical potential, and $\epsilon_{\bm{k},u}$ is the eigen energy of Eq.~\eqref{eq:WeylGeneral} with $u$ being the band index.
We obtain the expectation value of the Hund coupling $\ev{M\bm{\tau}}$ induced by the electric field $\dot{\bm{k}}=-e\bm{E}(t)/\hbar$ (i.e., magnetoelectric (ME) effect) as
\begin{align}
  \ev{M\bm{\tau}}     & =-\tau \frac{e}{\hbar}\frac{2\mu^2}{3(2\pi)^2|\det K|}M K\uprm{T}\bm{E}\nonumber            \\
                      & \equiv C\smrm{O} \bm{E}. \label{eq:MEWeyl}
\end{align}
Here, $C\smrm{O}$ is the linear ME tensor.
$C\smrm{O}$ is a dissipative term, which is proportional to the relaxation time $\tau$ and the square of the chemical potential $\mu^2$, i.e., the area of the Fermi surface.

Similarly, we obtain the current induced by fluctuations of the localized spins from the Boltzmann equation
$\ev{\bm{j}\smrm{O}} =C\smrm{O}\uprm{T}\dot{\bm{S}}(t)$, i.e., $\ev{\bm{j}\smrm{O}}(\omega) =-i\omega C\smrm{O}\uprm{T}\bm{S}(\omega)$.
This current response can be interpreted to arise from the emergent electric field due to the magnetization dynamics.

\subsection{Interband contribution to the ME effect: a two band model for a pair of Weyl fermions}
Next let us consideter the interband contribution to the ME effect induced by the Weyl fermions.
Motivated by the topological Fermi-sea response numerically demonstrated in Ref.~\cite{Meguro2025}, we consider an effective two-band model for a pair of Weyl points and derive an analytical expression for the interband contribution to the ME effect, which cannot be captured by the Boltzmann equation.
Specifically, we consider the Hamiltonian given by
\begin{align}
  \mathcal{H} & =(v k_x,v k_y,\frac{|\bm{k}|^2}{2m}-\eta)P\bm{\tau}+\bm{S}^\mathrm{T}(t)M\bm{\tau},\label{eq:compactWeylgeneral}   
\end{align}
where $v$ is the Fermi velocity, and $P$ is a three-by-three orthogonal matrix.
This model hosts two Weyl points located at $\bm{k}_\mathrm{w}=(0,0,\pm\sqrt{2m\eta})$. Unlike Eq.~\eqref{eq:WeylGeneral}, this system exhibits a magnetoelectric (ME) effect originating from interband contributions. Since the intraband contribution can be evaluated using Eq.~\eqref{eq:MEWeyl}, we here focus on the interband ME effect $C\smrm{I}$, where $\ev{M\bm{\tau}}=(C\smrm{O}+C\smrm{I})\bm{E}$.
Employing the Kubo formula for interband contributions, we obtain zero-th order terms in $\omega$ as
\begin{gather}
  C_{\mathrm{I}\lambda}^\mu  =2\hbar \mathrm{Im}\sum_{u\neq u'}\int \frac{d\bm{k}}{(2\pi)^3}\frac{\mel{u}{{j}_{\bm{k}}^\lambda}{u'}\mel{u'}{M^\mu\cdot \bm{\tau}}{u}}{(\xi_{\bm{k},u}-\xi_{\bm{k},u'})^2}f_0(\xi_{\bm{k},u})  \label{eq:CI}
\end{gather}
where $j_{\bm{k}}^v=-\frac{e}{\hbar}\partial_\lambda \mathcal{H}$ is the current velocity in the $v$-direction, $u,u'$ denote the eigenstates, and $M^\mu$ is the $\mu$-th row of the matrix $M$, regarded as a vector.

Assuming that the chemical potential lies at the Weyl points, we obtain ME tensor $C\smrm{I}$ as
\begin{gather}
  C_{\mathrm{I}}=\frac{1}{16\pi^2}e|P|MPQI,\label{eq:CIgivenByI}\\
                  Q\equiv\left(\begin{array}{ccc}
                    0 &-1& 0\\1&0&0\\0&0&0
                  \end{array}\right),\\\
  I\equiv \begin{cases}
    -\frac{4}{v}\sqrt{2m\eta}+\frac{4m^2v}{\sqrt{m^2v^2-2m\eta}}\arctan \sqrt{\frac{mv^2-2\eta}{2\eta }} & \eta\leq mv^2/2 \\
    -\frac{4}{v}\sqrt{2m\eta}+\frac{4m^2v}{\sqrt{-m^2v^2+2m\eta}} \tanh^{-1} \sqrt{\frac{-mv^2+2\eta}{2\eta}} & \eta>mv^2/2
  \end{cases}\label{eq:I}
\end{gather}
Although the expression is written in piecewise form, both cases are same when the definition of $\arctan$ is extended to the complex domain.

Dividing the Brillouin zone into regions with and without Fermi arcs along the $k_z$ direction (the arc length being ), the first term in Eq.~\eqref{eq:I} corresponds to the contribution from the region with Fermi arcs, while the second term arises from the region without them. 
In the regime of small Fermi velocity, $v < \sqrt{2\eta/m}$, the first term is dominant, which is essentially determined by the ratio of the Fermi arc length $2\sqrt{2m\eta}$ to the Fermi velocity $v$.
Since $P$ in Eq.~\eqref{eq:compactWeylgeneral} corresponds to $K$ in Eq.~\eqref{eq:WeylGeneral}, and $Q$ in Eq.~\eqref{eq:CIgivenByI} represents a $90^\circ$ rotation in the $xy$ plane, the interband contribution to the ME response described by $C\smrm{I}$ is perpendicular to the intraband contribution to the ME response described by $C\smrm{O}$.
The detailed derivations are presented in Appendix~\ref{sec:app:I}.
Similarly, we obatin the current response to fluctuations of the localized spins as $\ev{\bm{j}}=i\omega C\smrm{I}\uprm{T}\bm{S}(\omega)$.

\begin{figure}
  \centering
  \includegraphics[width=\linewidth]{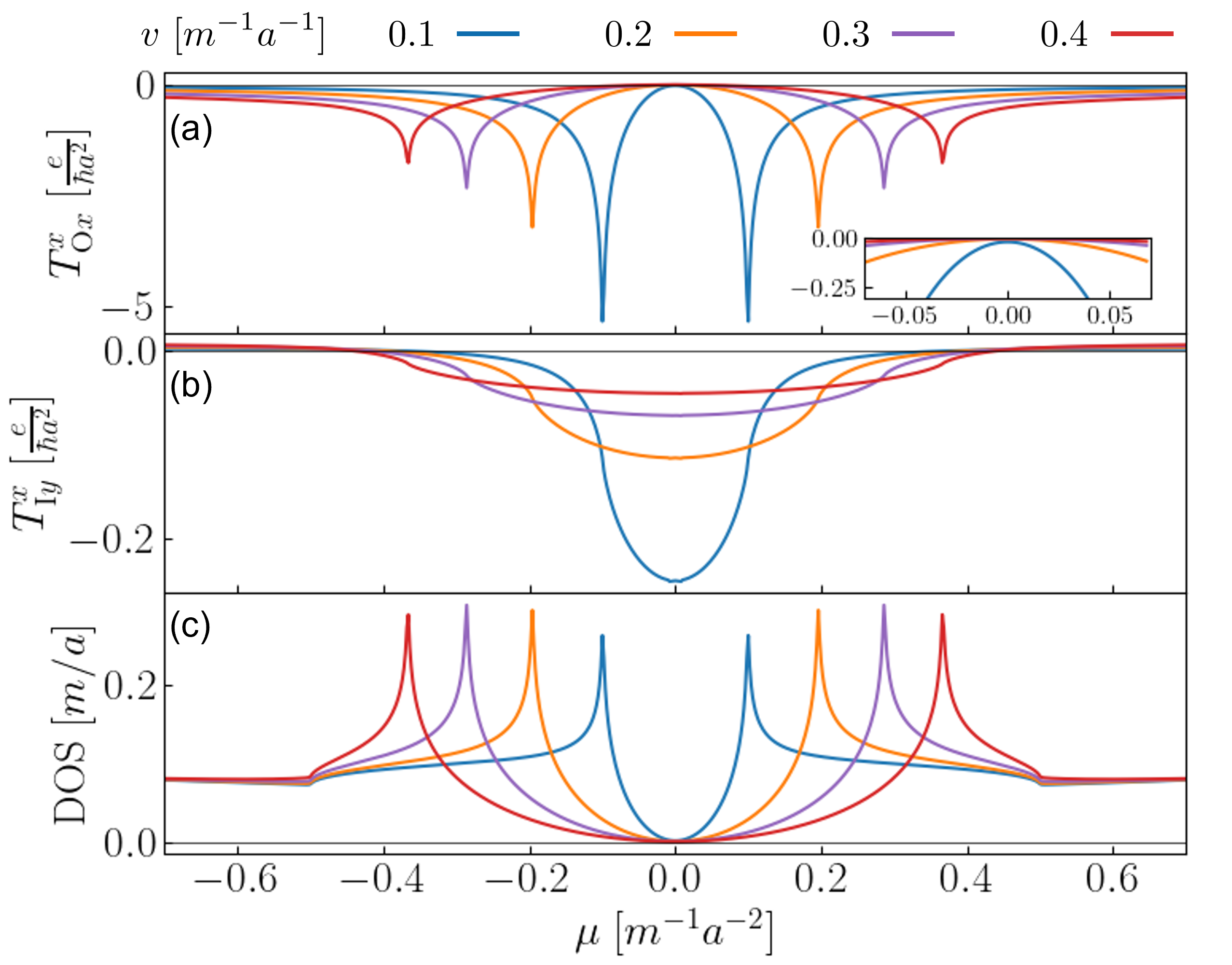}
  \caption{Chemical-potential dependence of (a) the Fermi surface contribution $C_{\mathrm{O}x}^x$ and (b) the interband contribution $C_{\mathrm{I}y}^x$ to the ME response, and (c) the density of states (DOS), calculated from the model in Eq.~\eqref{eq:compactWeylgeneral} for various values of the Fermi velocity $v$.
  $C_{\mathrm{O}x}^x$ is proportional to the square of the chemical potential, $\mu^2$, corresponding to the area of the Fermi surface in the Weyl-fermion region.
  $C_{\mathrm{I}y}^x$ emerges within the Fermi-velocity-induced gap, between the Van Hove singularities.
  We set the parameters as follows: $\tau=10ma^2\hbar$, $P$ is the identity matrix, and $M$ is a diagonal matrix whose diagonal elements are equal to $m^{-1}a^{-2}\hbar^{-1}$.}
  \label{fig:MEcompactWeyl}
\end{figure}

Figure~\ref{fig:MEcompactWeyl} shows the chemical potential dependence of the ME effect of (a) the Fermi surface term $C_{\mathrm{O}x}^x$ and (b) the interband effect term $C_{\mathrm{I}y}^x$ for the case $P$ being the identity matrix.
We set the parameters as $\tau=10ma^2\hbar$, $P$ is the identity matrix, and $M$ is a diagonal matrix whose diagonal elements are equal to $m^{-1}a^{-2}\hbar^{-1}$.
As shown in Fig.~\ref{fig:MEcompactWeyl}(b), the interband ME effect $C_{\mathrm{I}y}^x$ emerges within the Fermi velocity induced gap (between Van Hove singularities), and $C_{\mathrm{I}y}^x \propto v^{-1}$ at $\mu=0$.
As discussed in Sec.~\ref{sec:polarWeyl}, Eq.~\eqref{eq:compactWeylgeneral} can be regarded as the low-energy effective model of Eq.~\eqref{eq:minimalModel}. 
In this model, the Fermi velocity $v$ is proportional to the spin-orbit coupling strength, while the effective Hund coupling $M$ is proportional to the strength of the original Hund coupling. 
Accordingly, the Pauli matrices $\tau$ in Eq.~\eqref{eq:compactWeylgeneral} include spin degrees of freedom, ensuring that $M$ indeed represents a proper Hund coupling. 
Consequently, in the Weyl-fermion region, where the chemical potential lies between the Van Hove singularities, the interband contribution to the ME effect is inversely proportional to the spin-orbit coupling strength.

\section{Emergent induction}
In this section, we study emergent induction induced by Weyl fermions using the ME tensors obtained in the previous section.
In the emergent induction, the applied electric field induces a spin torque with the ME effect and the induced spin dynamics in turn gives rise to the emergent electric field (and thus electric current) again with the ME effect. 

To describe the emergent inductance due to the spin dynamics, we consider the impedance $Z$  defined by the inverse of the sum of conductivities as
\begin{align}
    Z=\frac{l}{A}(\sigma\smrm{DC}+\Sigma(\omega))^{-1},
\end{align}
where $l$ is the sample length, $A$ is the cross-sectional area, $\sigma\smrm{DC}$ is the DC conductivity, and $\Sigma(\omega)$ is the conductivity mediated by the dynamics of localized spins.
The conductivity mediated by the dynamics of localized spins is given by $\Sigma(\omega)=i\omega(-\tilde{C}\smrm{O} +\tilde{C}\smrm{I})^\mathrm{T}(-\omega\sigma_y-\mathcal{H}_m)^{-1}v(\tilde{C}\smrm{O}+\tilde{C}\smrm{I})$,
where $\mathcal{H}_m$is the coefficient matrix obtained by expressing the magnon Hamiltonian of localized spins in terms of
$S_x$ and $S_y$, and $v\smrm{cell}=a^3$ is the volume of the unit cell.
Here, $\tilde{C}\smrm{O(I)}$ is a two-by-three matrix obtained by removing the row corresponding to the $z$ component from $C\smrm{O(I)}$.
Therefore, the inductance $L\equiv -\mathrm{Im}[\partial_\omega Z]_{\omega=0}$ is given by
\begin{align}
  L=&\frac{lv\smrm{cell}}{A}\sigma\smrm{DC}^{-1}(-\tilde{C}\smrm{O} +\tilde{C}\smrm{I})^\mathrm{T}(-\mathcal{H}_m^{-1})(\tilde{C}\smrm{O}+\tilde{C}\smrm{I})\sigma\smrm{DC}^{-1}.
  \label{eq: L tensor}
\end{align}

From now, we consider the longitudinal (Hall) inductance, i.e., the symmetric (antisymmetric) part of the inductance,
\begin{align}
  L\smrm{long}=\frac{1}{2}(L+L^\mathrm{T}),\ L\smrm{Hall}=\frac{1}{2}(L-L^\mathrm{T}).
\end{align}
We can decompose the inductance into contributions from the Fermi surface and interband effects,$\sigma\smrm{DC}=\sigma\smrm{DC,O}+\sigma\smrm{DC,Hall}$,
and obtain the relations $\sigma\smrm{DC,O}=\sigma\smrm{DC,O}^\mathrm{T},\ \sigma\smrm{DC,Hall}=-\sigma\smrm{DC,Hall}^\mathrm{T}$.
We consider two cases: $\sigma\smrm{DC,O}\gg \sigma\smrm{DC,Hall},\sigma\smrm{DC}^{-1} \simeq \sigma\smrm{DC,O}^{-1}$ and $\sigma\smrm{DC,O}\ll \sigma\smrm{DC,Hall},\sigma\smrm{DC}\simeq \sigma\smrm{DC,I}$.
In both cases, the longitudinal and Hall inductances are given by
\begin{align}
  L\smrm{long}=&\frac{lv\smrm{cell}}{A}\sigma\smrm{DC}^{-1}[\tilde{C}\smrm{O}^\mathrm{T}\mathcal{H}_m^{-1}\tilde{C}\smrm{O}-\tilde{C}\smrm{I}^\mathrm{T}\mathcal{H}_m^{-1}\tilde{C}\smrm{I}]\sigma\smrm{DC}^{-1}\label{eq:LongInductance}\\
  L\smrm{Hall}=&\frac{lv\smrm{cell}}{A}\sigma\smrm{DC}^{-1}[\tilde{C}\smrm{O}^\mathrm{T}\mathcal{H}_m^{-1}\tilde{C}\smrm{I}-\tilde{C}\smrm{I}^\mathrm{T}\mathcal{H}_m^{-1}\tilde{C}\smrm{O}]\sigma\smrm{DC}^{-1}. \label{eq:HallInductance}
\end{align}
Here, we used the fact that $\mathcal{H}_m$ is a symmetric matrix.

\begin{figure}
  \centering
  \includegraphics[width=\linewidth]{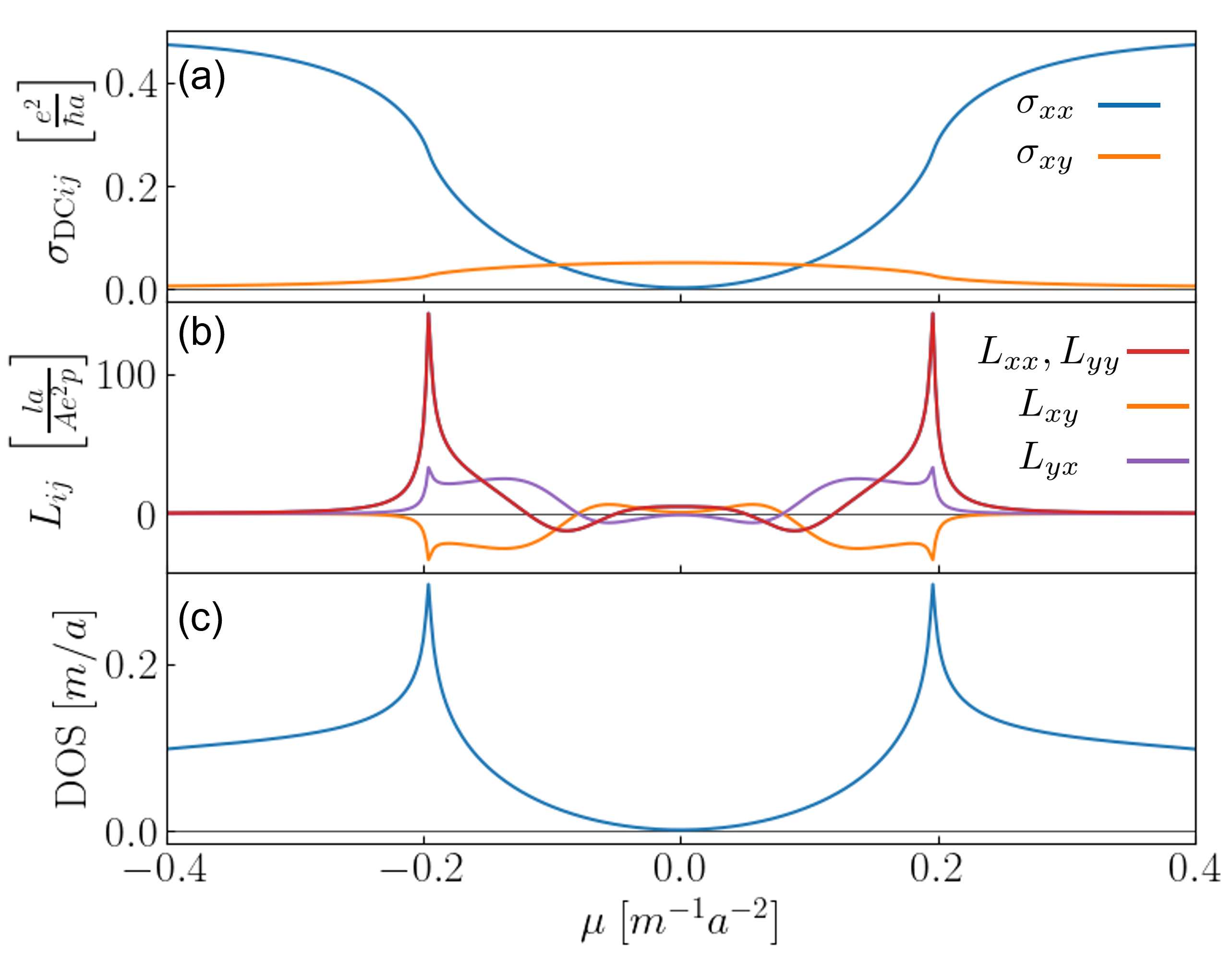}
  \caption{Chemical-potential dependence of (a) the conductivity and (b) the inductance, calculated from the model in Eq.~\eqref{eq:compactWeylgeneral} for $v=0.2/(ma)$.
  (c) The density of states (DOS) is replotted from Fig.~\ref{fig:MEcompactWeyl}(c) for comparison.
  The longitudinal conductivity $\sigma_{\mathrm{DC,O}xx}$ is proportional to $\mu^2$ in the Weyl-fermion region, while the Hall conductivity $\sigma_{\mathrm{DC,Hall}xy}$ is enhanced in the Weyl-fermion region.
  The inductances are also enhanced in the Weyl-fermion region.
  The parameters are the same as those in Fig.~\ref{fig:MEcompactWeyl}.}
  \label{fig:inductance}
\end{figure}

Figure~\ref{fig:inductance} shows the chemical potential dependence of the inductance calculated from the model in Eq.~\eqref{eq:compactWeylgeneral}.
We use the same parameters as those in Fig.~\ref{fig:MEcompactWeyl} and set the spin Hamiltonian of localized spins as $\mathcal{H}_m=pI_2$ with $I_2$ being the two-by-two identity matrix.
First, we focus on the longitudinal inductance.
In the region of $\sigma_{\mathrm{DC}xx}\ll \sigma_{\mathrm{DC}xy}$, $(|\mu|<0.05/(ma^2))$, $C_{\mathrm{O}_x}^x \ll C_{\mathrm{I}y}^x$, indicating that the contribution from the interband effect is dominant.
Since, $\mathcal{H}_m$ is a positive-definite symmetric matrix and $\sigma\smrm{DC}\simeq \sigma\smrm{DC,Hall}$ is antisymmetric, we obtain $L_{\mathrm{long}xx}>0$ by dropping $\tilde{C}\smrm{O}$ in Eq.~\eqref{eq:LongInductance}.
This shows that interband effects lead to a positive inductance, a phenomenon previously identified in emergent inductance at the surface of topological insulators~\cite{Araki2023}.
Our result shows that such positive inductance from interband effect also emerges in Weyl semimetals with suppressed longitudinal conductivity.
Notably, this situation is expected to exhibit a high quality factor, as reported in Ref.~\cite{Araki2023}.
On the other hand, in the region of $\sigma_{\mathrm{DC}xx}\gg \sigma_{\mathrm{DC}xy}$, $(0.15/(ma^2)<|\mu|<0.2)$, $C_{\mathrm{O}_x}^x \gg C_{\mathrm{I}y}^x$, indicating that the contribution from the Fermi surface effect is dominant.
In this region, since $\sigma\smrm{DC}\simeq \sigma\smrm{DC,O}$ is a symmetric matrix, we obtain $L_{\mathrm{long}xx}>0$.
This regime corresponds to the emergent inductance associated with the dynamics of classical localized spins, which has been widely studied~\cite{Nagaosa2019,Kurebayashi2021,Anan2025,Yamane2022}.

We next discuss the Hall inductance.
As shown in Eq.~\eqref{eq:HallInductance}, the Hall inductance is composed of terms that are linear in both $C\smrm{O}$ and $C\smrm{I}$.
For the case of $\sigma\smrm{DC,O}\gg \sigma\smrm{DC,Hall}$, the Hall inductance is proportional to $\tau^{-1}$ since $C\smrm{O}, \sigma\smrm{DC,O}\propto \tau$ and $C\smrm{I}$ is independent of $\tau$.
For the case of $\sigma\smrm{DC,O}\ll \sigma\smrm{DC,Hall}$, the Hall inductance is proportional to $\tau$.
The Hall inductance essentially originates from the interband effect $C\smrm{I}$ which captures a geometric nature of the Weyl semimetal. 

\section{Application to polar Weyl ferromagnets} \label{sec:polarWeyl}
In this section, we apply the formalism for the emergent induction of Weyl fermions to the case of polar Weyl ferromagnets.
A polar Weyl ferromagnet is a magnetic Weyl semimetal that lacks inversion symmetry and has polarity in the electronic states.
In insulators, polarity appears as an electric polarization $P$, which is screened in metals including Weyl semimetals.
When the electric polarization $P$ and the magnetization $M$ are present, one can generally define the toroidal moment $T=P\times M$.
The toroidal moment $T$ is a vector that is odd both under inversion and time reversal operation and behaves in the same manner as the electric current. 
In this case, the dynamics of magnetization is expected to induce current response in the direction of the toroidal moment $T$, and especially an emergent induction, which is termed ``emergent toroidal induction'' in Suzuki et al. ~\cite{Suzuki26}.
To study such emergent induction in Weyl semimetals, let us consider an effective model for a polar Weyl ferromagnet that breaks both inversion and time reversal symmetry.

Let us consider a model describing a polar Weyl ferromagnet which possesses a polarity $P$ and a magnetization $M$ along the $z$ direction.
Specifically, we study the Hamiltonian given by
\begin{align}
  \mathcal{H}=&\lambda \sin k_y a\tau_0 \sigma_x-\lambda \sin k_x a\tau_0 \sigma_y+\left(M_w+t_0\sum_\mu \cos k_\mu a \right)\tau_x \sigma_0  \nonumber \\
  &+\delta t \sin k_z a\tau_y\sigma_0+m_z(\tau_0+\alpha\tau_z)\sigma_z, \label{eq:minimalModel}
\end{align}
where $\sigma$ and $\tau$ are Pauli matrices representing spin and site degrees of freedom, respectively.
The first and second terms represent Rashba-type spin-orbit interactions, while the third term represents interorbital hopping.
The third term is the mass term and
the fourth term represents the staggered hopping along the $z$ direction.
The fifth term represents the Hund coupling between itinerant electrons and localized spins, where $m_z$ is the magnitude of the Hund coupling and $\alpha$ is a parameter that characterizes the difference in the Hund coupling between the two orbitals.
The fourth and fifth terms break both inversion symmetry and time-reversal symmetry, which is a characteristic feature of polar Weyl ferromagnets. 
The fourth and fifth terms break inversion symmetry and introduce a polarity along the $z$ direction.
Combined with time-reversal symmetry breaking due to the fifth term, this model captures a characteristic feature of polar Weyl ferromagnets.
The model has 4-fold rotation symmetry along the $z$ axis, 
and also preserves the mirror symmetry along the $x$ and $y$ directions with $R_x=i\sigma_x$ and $R_y=i\sigma_y$ in the case of $m_z=0$.

Expanding the Hamiltonian near the $\Gamma$ point $(0,0,0)$, we obtain  effective Hamiltonians of the form given in Eq.~\eqref{eq:compactWeylgeneral}.
The parameters in Eq.~\eqref{eq:compactWeylgeneral} are given as
\begin{gather}
  v=\lambda a  \cos\theta,\\
  m=t_0^{-1}a^{-2}/\sin\theta,\\
  \eta=\sqrt{(M_w+3t_0)^2 + m_z^2 \alpha^2}-m_z,\\
  P=\begin{pmatrix}
     0&- 1& 0\\
    -1&0&  0\\
    0& 0& -1
  \end{pmatrix},\\
  M=\begin{pmatrix}
    (\alpha-\cos \theta)& 0& 0\\
    0& -(\alpha-\cos \theta)& 0\\
    0& 0& -1+\alpha \cos\theta
  \end{pmatrix},
\end{gather}
where $\theta$ is defined as $\tan \theta = 3t_0/(m_z \alpha)$.
These expressions provide a physical interpretation of the parameters in Eq.~\eqref{eq:compactWeylgeneral}.
The effective Fermi velocity $v$ is regarded as the renormalized spin-orbit coupling, which is proportional to the inversion-symmetry-breaking parameter $\alpha$.
The mass term $m$ stems from the hopping parameter $t_0$ in the form $m^{-1}a^{-2} \propto t_0$, and $\eta$ is determined by the magnetization $m_z$ and the inversion-symmetry-breaking parameter $\alpha$.
The effective Hund coupling with $S_x(t)$ and $S_y(t)$ is proprortional to the inversion-symmetry-breaking parameter $\alpha$.
The details of the derivation are presented in Appendix~\ref{app:polarWeyl}.

\begin{figure}
  \centering
  \includegraphics[width=\linewidth]{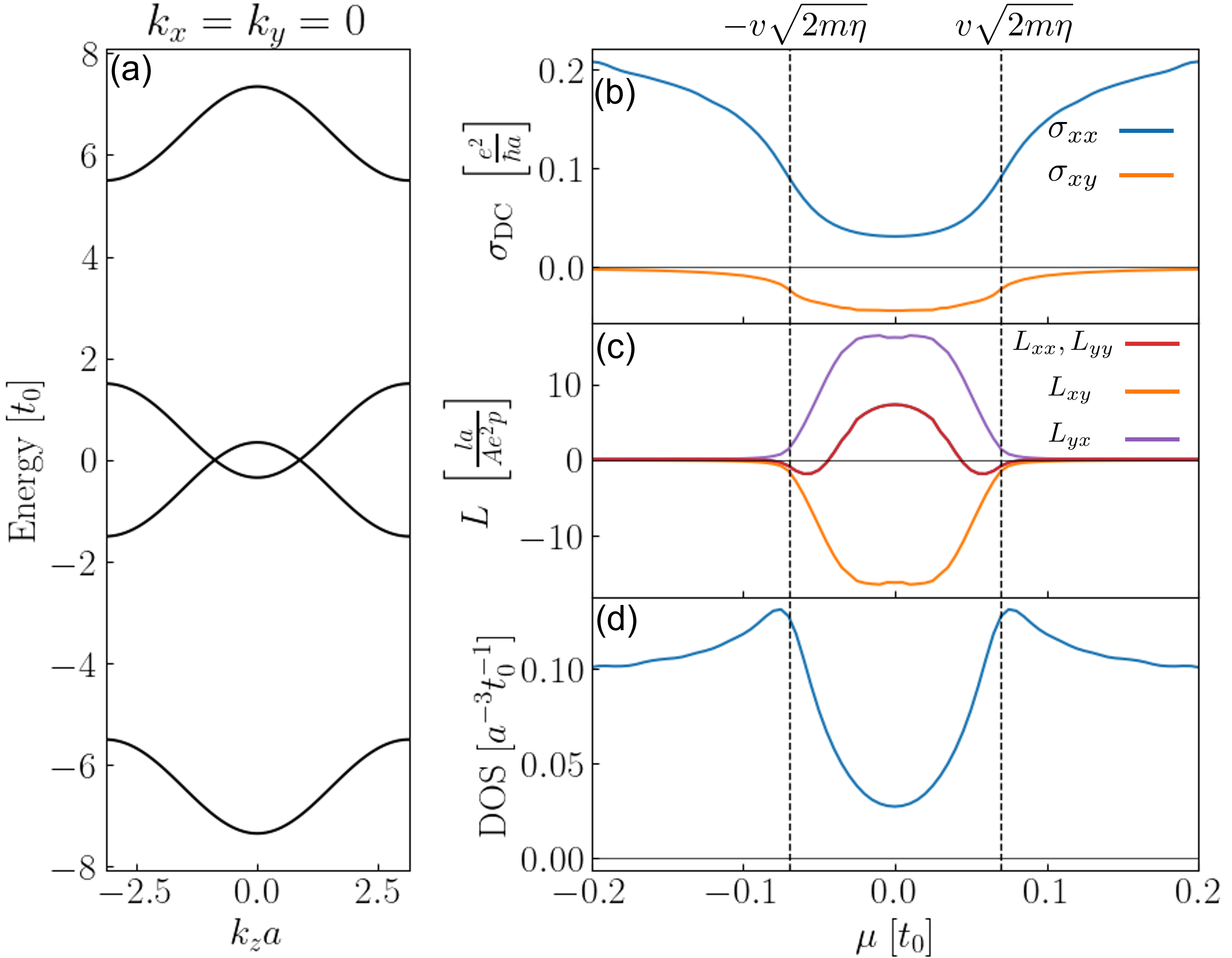}
  \caption{(a) Band structure and (b) the conductivity and (c) the inductance obtained from the model in Eq.~\eqref{eq:minimalModel}. 
  The dashed lines in (b-d) indicate the energy range of Weyl dispersion $|\mu| < v\sqrt{2m\eta}$.
  Similar to the results in Fig.~\ref{fig:inductance}, the inductance are enhanced in the Weyl-fermion region, $|\mu|<v\sqrt{2m\eta}$.
  The parameters are $\tau=10\hbar/t_0$, $\lambda =0.3t_0$, $M_w=0.7t_0$, $\delta t=0.1t_0$, $m_z=3.5t_0$, and $\alpha=0.3$}
  \label{fig:polarWeyl}
\end{figure}

Figure~\ref{fig:polarWeyl} shows the band structure and the chemical potential dependence of the conductivity and the inductance obtained from the model in Eq.~\eqref{eq:minimalModel}.
We set the parameters to $\tau=10\hbar/t_0$, $\lambda=0.3t_0$, $M_w=0.7t_0$, $\delta t=0.1t_0$, $m_z=3.5t_0$, and $\alpha=0.3$, and take the spin Hamiltonian for the localized spins to be $\mathcal{H}_m=pI_4$, where $I_4$ is the four-by-four identity matrix.
Figure~\ref{fig:polarWeyl}(a) shows that the model has two Weyl points located at $(0,0,\pm \sqrt{2m\eta}/a)=(0,0,\pm 0.8/a)$.
With these parameters, the effective Fermi velocity is $v=\lambda a \cos\theta=0.08a t_0$ and the range that the Hall conductance $\sigma_{xy}$ and the inductance are enhanced is $|\mu|<v\sqrt{2m\eta}$ as shown in Fig.~\ref{fig:polarWeyl}(b,c).
This is consistent with the result obtained from the effective model in Eq.~\eqref{eq:compactWeylgeneral}.

There are several contributions to the emergent inductance in the general formula given in Eq.~\eqref{eq: L tensor}. 
In the typical case where the Hall angle is small, i.e., $\sigma\smrm{DC,O}\gg \sigma\smrm{DC,Hall}$, the inductance reduces to the forms given in Eqs.~\eqref{eq:LongInductance} and \eqref{eq:HallInductance}. Combinations of intraband ME tensors $C\smrm{O}$ or interband ME tensors $C\smrm{I}$ give rise to a longitudinal inductance, whereas combinations of intraband and interband ME tensors lead to a Hall inductance.
In particular, one contribution to the Hall induction is given as follows. 
By applying the electric field along the $x$ direction, the interband ME coupling $C\smrm{I}$ induces the magnetization along the $x$ direction ($\delta M_x $).
The induced magnetization $\delta M_x$ gives rise to the current response along the $y$ direction ($J_y$) through the intraband ME coupling $C\smrm{O}$.
Since the intraband ME coupling $C\smrm{O}$ describes the emergent electric field due to the Weyl point shift, this current response can be interpreted that $J_y$ is induced by the emergent electric field $e_y$ along the $y$ direction due to the magnetization dynamics.
Since the system has polarity along the $z$ direction, the direction of the emergent electric field coincides with that of the toroidal moment $T=P \times M$ with assuming the screened polarization $P \parallel z$.
Note that the other term for the Hall inductance in Eq.~\eqref{eq:HallInductance} corresponds to the process where the magnetization along $y$ is induced by the Fermi surface shift under the Rashba-type spin-momentum locking and induced emergent electric field along the $x$ gives rise to a Hall current along the $y$ direction.
The first term in Eq.~\eqref{eq:HallInductance} corresponds to the contribution regarded as toroidal induction.

\section{Discussion}
We have shown that the magnetic Weyl semimentals generally serve as an emergent inductor.
The inductive response originates from the magnetoelectric coupling for Weyl fermions, which has intraband and interband contributions.
The intraband contribution of the ME coupling corresponds to the emergent electric field arising from the shift of the Weyl point in the momentum space.

An emergent inductive response is generally expected when the two-dimensional subspace of the Weyl fermionic states contains the spin degrees of freedom. 
In the case of polar Weyl ferromagnets with Rashba spin-orbit coupling, the Hall inductance appears along the direction of the toroidal moment induced by the magnetization dynamics in addition to the longitudinal inductance.
This follows from the general analysis of the magnetoelectric response of Weyl fermions, indicating that the underlying symmetry of the  Weyl semimetal determines the directionality of the inductive response.
As another example, Weyl fermions emerging at the phase boundary between three-dimensional topological and trivial insulators \cite{Murakami07,Halasz12} are also expected to show emergent inductance once they are coupled to magnetization as in magnetic topological insulators~\cite{Tokura2019}. 
In those cases, the low-energy effective Weyl Hamiltonian generally cannot be reduced to the simple form $\bm{m}\cdot\bm{\sigma}$ with the spin Pauli matrices $\bm{\sigma}$. Consequently, the direction and anisotropy of the emergent inductance depend on the microscopic structure of the Weyl Hamiltonian.

Weyl semimetals exhibit an enhancement of the emergent inductance when the chemical potential lies in the Weyl fermion regime. 
This enhancement originates from the interband contribution to the magnetoelectric response, which is amplified by the small energy separation between the two Weyl bands. 
Since this low-energy band structure is an intrinsic property of Weyl fermions, the enhancement mechanism is expected to be robust and does not depend on details of the microscopic Hamiltonian.

The present study suggests that Weyl semimetals provide a suitable platform for investigating emergent electromagnetic responses. The enhancement of the emergent inductance, originating from the low-energy structure of Weyl fermions, illustrates how topological band structures can give rise to emergent electromagnetic responses. It is an interesting future issue to study whether similar mechanisms arise in other topological semimetal systems.

\begin{acknowledgments}
We thank Yukako Fujishiro, Yuuri Suzuki, Masataka Mogi, Manabu Sato, Hiroaki Ishizuka, Yasufumi Araki and Yoshinori Tokura for inspiring discussions. This work was supported by JST SPRING, Grant Number JPMJSP2108 (TA) and MEXT/JSPS KAKENHI, Grants Numbers JP26KJ0822 (TA), JP24H02231, JP24K00568, JP23K25816 (TM).
\end{acknowledgments}

\appendix
\section{Derivation of $I$}\label{sec:app:I}
In this appendix, we derive Eq.~\eqref{eq:I} from Eq.~\eqref{eq:CI}.
By defining $\bm{R}$ as $\mathcal{H}=\bm{R}\cdot \bm{\sigma}+\bm{S}^{\mathrm{T}}M\bm{\tau}$ with $\mathcal{H}$ being the Hamiltonian shown in Eq.~\eqref{eq:compactWeylgeneral}, we obtain
\begin{align}
    C_{\mathrm{I}\mu}^\lambda=\frac{e}{2}\int \frac{d\bm{k}}{(2\pi)^3}\frac{M^\mu\cdot (\bm{R}\times \partial_\lambda \bm{R})}{|\bm{R}|^3}
\end{align}
We obtain
\begin{align}
  |\bm{R}|=                       & \sqrt{v^2 k_x^2+v^2 k_y^2+\left(\frac{|\bm{k}|^2}{2m}-\eta\right)^2}\nonumber                                                 \\
  =                               & \sqrt{v^2 r^2+\frac{1}{4m^2}(r^2+k_z^2-2m\eta)^2},\label{eq:R}  \\
\bm{R}\times \partial_x \bm{R}=& |P|P\left(\begin{array}{c}
  \frac{v}{2m}r^2\sin 2\phi\\ -\eta v +\frac{v}{2m}k_z^2-\frac{v}{2m}r^2\cos 2\phi\\ -v^2 r \sin \phi
  \end{array}\right),\\
\bm{R}\times \partial_y \bm{R}=& |P|P\left(\begin{array}{c}
  \eta v -\frac{v}{2m}k_z^2-\frac{v}{2m}r^2\cos 2\phi\\ -\frac{v}{2m}r^2\sin 2\phi\\ -v^2 r \cos \phi
  \end{array}\right),\\
\bm{R}\times \partial_z \bm{R}=& |P|P\left(\begin{array}{c}
  \frac{v}{m}k_z r\sin\phi \\ -\frac{v}{m}k_z r\cos\phi \\ 0
  \end{array}\right).
\end{align}
Here, cylindrical coordinates have been introduced.
Therefore, the magnetization response is given by
\begin{align}
  \frac{\pi}{2}e|P|MP\left(\begin{array}{ccc}
                    0 &-1& 0\\1&0&0\\0&0&0
                  \end{array}\right)\int\frac{dk_z rdr}{(2\pi)^3}2\frac{-\eta v +\frac{v}{2m}k_z^2}{\left[v^2 r^2+\frac{1}{4m^2}(r^2+k_z^2-2m\eta)^2\right]^{3/2}} .
\end{align}
Below, we evaluate
\begin{align}
  I\equiv \int dk_z rdr\,2\frac{-\eta v +\frac{v}{2m}k_z^2}{\left[v^2 r^2+\frac{1}{4m^2}(r^2+k_z^2-2m\eta)^2\right]^{3/2}} .
\end{align}
By setting $z=r^2$, we obtain
\begin{align}
  I= & \int dk_z dz\frac{-\eta v +\frac{v}{2m}k_z^2}{\left[v^2 z+\frac{1}{4m^2}(z+k_z^2-2m\eta)^2\right]^{3/2}}\nonumber            \\
  =  & \int \frac{dk_z dz\ 4m^2v(-2m\eta+k_z^2)}{\left[(z+k_z^2-2m\eta+2m^2v^2)^2-4m^2v^2(k_z^2-2m\eta +m^2v^2)\right]^{3/2}} .
\end{align}

\subsection{Case of $\eta\leq mv^2/2$}
First, we consider the case of $\eta\leq mv^2/2$.
We introduce
\begin{align}
  x & \equiv z+k_z^2-2m\eta+2m^2v^2,              \\
  b & \equiv 2mv\sqrt{k_z^2-2m\eta +m^2v^2}.
\end{align}
Then we can write
\begin{align}
  I= & \int_{-\infty}^\infty dk_z \int_{k_z^2+2m^2v^2-2m\eta}^{\infty} dx\frac{4m^2v(-2m\eta+k_z^2)}{(x^2-b^2)^{3/2}} .
\end{align}
Setting $x=b/\cos\theta$, we obtain
\begin{align}
  I= & \int_{-\infty}^\infty dk_z \frac{4m^2v(-2m\eta +k_z^2)}{b^2}\int_{\theta_0}^{\frac{\pi}{2}} d\theta \frac{\cos\theta}{\sin^2\theta}\nonumber \\
  =  & \int_{-\infty}^\infty dk_z \frac{4m^2v(-2m\eta +k_z^2)}{b^2}\left[-\frac{1}{\sin\theta}\right]_{\theta_0}^{\frac{\pi}{2}},
\end{align}
with
\begin{align}
  \cos\theta_0= & \frac{b}{k_z^2+2m^2v^2-2m\eta},                                         \\
  \sin\theta_0= & \frac{\sqrt{(k_z^2+2m^2v^2-2m\eta)^2-b^2}}{k_z^2+2m^2v^2-2m\eta} \nonumber\\
  =             & \frac{|k_z^2-2m\eta|}{k_z^2+2m^2v^2-2m\eta}.
\end{align}
Since $k_z^2+2m^2v^2-2m\eta>0$, both $\cos\theta_0$ and $\sin\theta_0$ are positive, which implies $0<\theta_0<\pi/2$.
Thus,
\begin{align}
  I=\int_{-\infty}^\infty dk_z \frac{4m^2v(-2m\eta +k_z^2)}{b^2}\left[-1+\frac{k_z^2+2m^2v^2-2m\eta}{{|k_z^2-2m\eta|}}\right].
\end{align}
Although the expression after the $z$ integration contains $|k_z^2-2m\eta|$ in the denominator, this does not lead to a
nonintegrable singularity. 
Indeed, with $q=k_z^2-2m\eta$, the $z$-integrated integrand has the finite
one-sided limits
\begin{align}
\lim_{q\to 0\pm}\frac{4m^2v(-2m\eta +k_z^2)}{b^2}\left[-1+\frac{k_z^2+2m^2v^2-2m\eta}{{|k_z^2-2m\eta|}}\right]=\pm\frac{2}{v}
\end{align}
Thus $q=0$ gives only a finite jump discontinuity. 
Since this occurs only at the isolated points $k_z=\pm\sqrt{2m\eta}$, which have zero measure in the $k_z$ integration, their values do not affect $I$.
We therefore evaluate the integral on the open intervals separated by these points and omit the endpoints.

For $|k_z|<\sqrt{2m\eta}$, this gives
\begin{align}
  I_{|k_z|<\sqrt{2m\eta}}= & \int_{-\sqrt{2m\eta}}^{\sqrt{2m\eta}} dk_z \left(-\frac{2}{v}\right)\nonumber \\
  =                        & -\frac{4}{v}\sqrt{2m\eta}.
\end{align}
For $|k_z|>\sqrt{2m\eta}$, we obtain
\begin{align}
  I_{|k_z|>\sqrt{2m\eta}}= & \left[\int_{-\infty}^{\infty}-\int_{-\sqrt{2m\eta}}^{\sqrt{2m\eta}} \right] dk_z \frac{2m^2v}{k_z^2+m^2v^2-2m\eta}.
\end{align}
With the substitution $k_z=\sqrt{m^2v^2-2m\eta}\tan\theta$, the integral is evaluated as
\begin{align}
  I_{|k_z|>\sqrt{2m\eta}}= & \left[\pi-2\arctan \sqrt{\frac{2\eta }{mv^2-2\eta}}\right]\frac{2m^2v}{\sqrt{m^2v^2-2m\eta}} \nonumber\\
  =                        & \frac{4m^2v}{\sqrt{m^2v^2-2m\eta}}\arctan \sqrt{\frac{mv^2-2\eta}{2\eta}}.
\end{align}
Here, we have used $\pi/2=\arctan x+\arctan (1/x)$.
Therefore,
\begin{align}
  I= & -\frac{4}{v}\sqrt{2m\eta}+\frac{4m^2v}{\sqrt{m^2v^2-2m\eta}}\arctan \sqrt{\frac{mv^2-2\eta}{2\eta}} .
\end{align}

\subsection{Case of $\eta>mv^2/2$}
For $|k_z|\geq \sqrt{-m^2v^2+2m\eta}$, the $z$ integral can be performed in the same way as in the previous subsection:
\begin{align}
    & I_{|k_z|\geq \sqrt{-m^2v^2+2m\eta}}\nonumber                                                                                              \\
  = & \left[\int_{-\infty}^{\infty} -\int_{-\sqrt{-m^2v^2+2m\eta}}^{\sqrt{-m^2v^2+2m\eta}}\right]dk_z \frac{4m^2v(-2m\eta +k_z^2)}{b^2}\nonumber \\
    & \quad \times \left[-1+\frac{k_z^2+2m^2v^2-2m\eta}{{|k_z^2-2m\eta|}}\right].
\end{align}
Hence, we can write
\begin{align}
  I_{|k_z|>\sqrt{2m\eta}}= & 2\int_{\sqrt{2m\eta}}^\infty dk_z \frac{2m^2v}{k_z^2+m^2v^2-2m\eta}\nonumber                                                        \\
  =                        & \frac{2m^2v}{\sqrt{-m^2v^2+2m\eta}}\log \frac{\sqrt{2m\eta}+\sqrt{-m^2v^2+2m\eta}}{\sqrt{2m\eta}-\sqrt{-m^2v^2+2m\eta}}\nonumber \\
  =&\frac{4m^2v}{\sqrt{-m^2v^2+2m\eta}} \tanh^{-1} \sqrt{\frac{-mv^2+2\eta}{2\eta}} .
\end{align}
For $\sqrt{-m^2v^2+2m\eta}<|k_z|<\sqrt{2m\eta}$, we obtain
\begin{align}
  I_{\sqrt{-m^2v^2+2m\eta}<|k_z|<\sqrt{2m\eta}}= & \int_{\sqrt{-m^2v^2+2m\eta}<|k_z|<\sqrt{2m\eta}} dk_z \left(-\frac{2}{v}\right).
\end{align}

For $|k_z|<\sqrt{-m^2v^2+2m\eta}$, by defining $c=2mv\sqrt{-k_z^2-m^2v^2+2m\eta}$, we obtain
\begin{align}
  &I_{|k_z|<\sqrt{-m^2v^2+2m\eta}}\nonumber \\
  = & \int_{-\sqrt{-m^2v^2+2m\eta}}^{\sqrt{-m^2v^2+2m\eta} } dk_z \int_{k_z^2+2m^2v^2-2m\eta}^\infty dx\frac{4m^2v(-2m\eta+k_z^2)}{(x^2+c^2)^{3/2}},
\end{align}
where $x\equiv z+k_z^2-2m\eta+2m^2v^2$.
Using $x=c\tan\theta$, we find
\begin{align}
    & I_{|k_z|<\sqrt{-m^2v^2+2m\eta}}\nonumber                                                                                                                           \\
  = & \int_{-\sqrt{-m^2v^2+2m\eta}}^{\sqrt{-m^2v^2+2m\eta} }dk_z \frac{4m^2v(-2m\eta +k_z^2)}{c^2}\int_{\theta_0}^{\frac{\pi}{2}} d\theta \cos\theta \nonumber \\
  = & \int_{-\sqrt{-m^2v^2+2m\eta}}^{\sqrt{-m^2v^2+2m\eta} }dk_z \left(-\frac{2}{v}\right).
\end{align}
Here, the sign of $\sin\theta_0$ depends on the sign of $k_z^2-2m\eta+2m^2v^2$.
Combining these contributions gives
\begin{align}
  I= & -\frac{4}{v}\sqrt{2m\eta}+\frac{4m^2v}{\sqrt{-m^2v^2+2m\eta}} \tanh^{-1} \sqrt{\frac{-mv^2+2\eta}{2\eta}} .
\end{align}

\section{Parameter transformation for the minimal model}\label{app:polarWeyl}
The minimal model in Eq.~\eqref{eq:minimalModel} contains orbital degrees of freedom $\tau$ and spin degrees of freedom $\sigma$.
This model can host Weyl points. 
For $m_z=3.5t_0$, $\alpha=0.3$, $M_w=0.7t_0$, and $\delta t=0.1t_0$, we find that the Weyl points are located at $(0,0,\pm 0.8/a)$ and the Fermi arc passes through the $\Gamma$ point, $(0,0,0)$.
The effective Hamiltonian near the $\Gamma$ point is derived by performing a unitary transformation at the $\Gamma$ point and expanding the Hamiltonian in powers of $k_x$, $k_y$, and $k_z$.
These are given by
\begin{align}
  \mathcal{H}=&-\lambda \cos \theta ( k_y a\sigma_x +  k_x a\sigma_y) \nonumber\\
  &+ \{-m_z + [M_w+t_0(3-\frac{1}{2}|\bm{ k}|^2a^2)] \sin\theta + m_z \alpha \cos\theta \}\sigma_z,
\end{align}
where $\cos\theta = \frac{m_z \alpha}{\sqrt{(M_w+3t_0)^2 + m_z^2 \alpha^2}}$ and $\sin \theta=\frac{M_w+3t_0}{\sqrt{(M_w+3t_0)^2 + m_z^2 \alpha^2}}$.
We also obtain the effective Hund coupling by performing the same unitary transformation to $\bm{S}(t)\cdot (\tau_0+\alpha\tau_z)\bm{\sigma}$, which is given by
\begin{align}
  (\alpha-\cos \theta)[ S_x(t)\sigma_x -S_y(t)\sigma_y ]+(-1+\alpha \cos\theta)S_z (t).
\end{align}
Therefore, the parameters in Eq.~\eqref{eq:compactWeylgeneral} are given as
\begin{gather}
  v=\lambda a  \cos\theta,\\
  m=t_0^{-1}a^{-2}/\sin\theta,\\
  \eta=\sqrt{(M_w+3t_0)^2 + m_z^2 \alpha^2}-m_z,\\
  P=\begin{pmatrix}
     0&- 1& 0\\
    -1&0&  0\\
    0& 0& -1
  \end{pmatrix},\\
  M=\begin{pmatrix}
    (\alpha-\cos \theta)& 0& 0\\
    0& -(\alpha-\cos \theta)& 0\\
    0& 0& -1+\alpha \cos\theta
  \end{pmatrix}
\end{gather}
Therefore, in Eq.~\eqref{eq:compactWeylgeneral}, the effective Fermi velocity $v$ is regarded as the renormalized spin-orbit coupling, which is proportional to the inversion-symmetry-breaking parameter $\alpha$.
$m^{-1}a^{-2}$ stem from the hopping parameter $t_0$ and $\eta$ is determined by the magnetization $m_z$ and the inversion-symmetry-breaking parameter $\alpha$.
The effective Hund coupling with $S_x(t)$ and $S_y(t)$ is proprortional to the inversion-symmetry-breaking parameter $\alpha$.

\bibliography{ref}
\end{document}